\documentclass[aps,prl,floatfix,twocolumn]{revtex4}
\usepackage{graphics}
\usepackage{dcolumn}
\usepackage{epsfig}

\newcommand{\bk}{{\bf k }}

\newcommand{\bq}{{\bf q }}

\newcommand{\EP}{{\it e}-ph}

\begin{document}

\title
{Velocity Renormalization and Carrier Lifetime in Graphene
\\ from Electron-Phonon Interaction}
\author{Cheol-Hwan Park}
\email{cheolwhan@civet.berkeley.edu}
\author{Feliciano Giustino}
\author{Marvin L. Cohen}
\author{Steven G. Louie}
\affiliation{Department of Physics, University of California at Berkeley,
Berkeley, California 94720\\
Materials Sciences Division, Lawrence Berkeley National
Laboratory, Berkeley, California 94720}

\date{\today}

\begin{abstract}
We present a first-principles investigation of the phonon-induced electron self-energy
in graphene.
The energy dependence of the self-energy reflects the peculiar linear bandstructure
of graphene and deviates substantially from the usual metallic 
behavior. The effective band velocity of the Dirac fermions is found to be reduced by 
4-8~\%, depending on doping, by the interaction with lattice vibrations. 
Our results are consistent with the observed
linear dependence of the electronic linewidth on the binding energy
in photoemission spectra.
\end{abstract}
\maketitle

The recent fabrication of single-layer graphene~\cite{novoselov:2005PNAS_2D} has
attracted considerable interest because low-energy charge carriers in this material
have dispersion curves similar to Dirac fermions with zero rest mass and constant group 
velocity~\cite{novoselov:2005Nat_Graphene_QHE,zhang:2005Nat_Graphene_QHE}.
Because of the peculiar electronic structure of graphene, electrons and holes exhibit
exceptionally large mobilities, and the density of states can be tuned over a wide range
by applying a gate voltage~\cite{novoselov:2005Nat_Graphene_QHE, zhang:2005Nat_Graphene_QHE}.
These properties make graphene a promising candidate for new-generation
electronic and spintronic devices.

Angle-resolved photoemission spectroscopy (ARPES) is used as a powerful
tool for investigating quasiparticle behavior with extremely fine energy
and momentum resolution~\cite{damascelli:2003RMP_ARPES}.
The photoelectron intensity
provides information about the energy vs.\ momentum dispersions of
the charge carriers and the associated lifetimes. Recent photoemission experiments performed
on graphene showed a peculiar dependence of the hole lifetime on the binding energy, as well
as a significant velocity renormalization~\cite{bostwick:2007NatPhys_Graphene}.
The measured carrier lifetime has been discussed
within a model including three different decay channels: electron-phonon (\EP) scattering,
electron-plasmon scattering, and electron-hole pair
generation~\cite{bostwick:2007NatPhys_Graphene}.
The linear dependence of the linewidth on the binding energy was
attributed to the generation of electron-hole pairs.
The phonon-induced lifetime was assumed to be
energy-independent 
as found in conventional metallic systems~\cite{bostwick:2007NatPhys_Graphene}.
A subsequent theoretical work analyzed the carrier lifetimes in graphene
by adopting a two-dimensional electron-gas model,
and concluded that the experimental results could be explained without
invoking the \EP\ interaction~\cite{hwang:2006condmat_Graphene_GW}.

In this work we investigate the \EP\ interaction in graphene
within a {\it first-principles} approach.
We calculated the
electron self-energy arising from the \EP\ interaction
using a dense sampling of
the scattering processes in momentum space, and we extracted
the velocity renormalization and the carrier lifetimes from the corresponding
real and imaginary parts of the self-energy, respectively.
Our analysis shows that the self-energy associated with the \EP\ interaction 
in graphene is qualitatively different from that found in conventional metals.
The imaginary part of the self-energy shows a linear energy dependence
above the phonon emission threshold,
which directly reflects the bandstructure of graphene.
The real part of the self-energy leads to a
Fermi velocity renormalization of 4-8~\% depending on doping.
We further propose a simple analytical model of the electron self-energy capturing 
the main features of our first-principles calculations. 
Our calculation allows us to assign the low-energy kink in the measured photoemission
spectrum and part of the linear energy-dependence of the electronic linewidths
to the \EP\ interaction. 

The \EP\ interaction in graphene is treated within
the Migdal approximation~\cite{grimvall:1981_Metal_ElPh}.
The contribution to 
the electron self-energy $\Sigma_{n\bk}(E;T)$ arising from the \EP\ interaction 
at the temperature $T$ is~\cite{grimvall:1981_Metal_ElPh, giustino:2007TBP_Migdal}:
  \begin{eqnarray}
  &&\Sigma_{n\bk}(E;T)  =  \sum_{m,\nu} \, \int \, \frac{d\bq}{A_{\rm BZ}}
  \; |g_{mn,\nu}(\bk,\bq)|^2 \nonumber \\
  && \hspace{-0.5cm}\times \left[\frac{n_{\bq\nu}
  +1-f_{m\bk+\bq}}
  {E-\varepsilon_{m\bk+\bq}-\hbar\omega_{\bq\nu}-i\delta}
  +\frac{n_{\bq\nu}+f_{m\bk+\bq}}
  {E-\varepsilon_{m\bk+\bq}+\hbar\omega_{\bq\nu}-i\delta}
  \right], \nonumber \\ 
  \label{equation:selfenergy}
  \end{eqnarray}
where $\varepsilon_{n\bk}$ is the energy of an electronic state with band index $n$
and wavevector $\bk$, and $\hbar\omega_{\bq\nu}$
the energy of a phonon with wavevector $\bq$ and branch index $\nu$. 
$f_{n\bk}$ and $n_{\bq\nu}$, are the Fermi-Dirac
and Bose-Einstein distribution functions,
respectively. The integration
extends over the Brillouin zone (BZ) of graphene of area $A_{\rm BZ}$
and the sum runs over
both occupied and empty electronic states
and all phonon branches.
The \EP\ matrix element is defined by
$g_{mn,\nu}(\bk,\bq)= \langle m\bk+\bq|\Delta V_{\bq\nu}|n\bk \rangle$,
$\Delta V_{\bq\nu}$ being the change in the self-consistent potential due to a
phonon with wavevector $\bq$ and branch index $\nu$,
while $|n\bk \rangle$, $|m\bk+\bq \rangle$  indicate Bloch eigenstates.
Equation (\ref{equation:selfenergy}) takes into account the anisotropy of the
\EP\ interaction in $\bk$-space, as well as retardation effects 
through the phonon frequency in the denominators.

\begin{figure}
  \includegraphics[width=1.0\columnwidth]{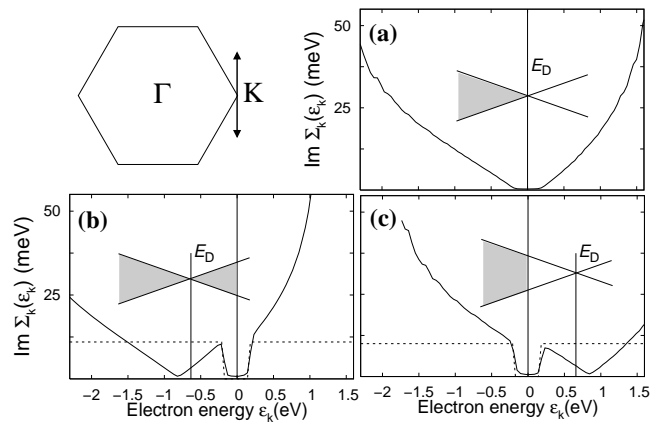}
  \caption{Calculated imaginary part of the electron self-energy arising from
  the \EP\ interaction at $T=$ 20~K (solid lines), for (a) intrinsic,
  (b) electron-doped, and (c) hole-doped graphene.
  The self-energy $\Sigma_\bk(\varepsilon_\bk)$
  was evaluated along the reciprocal space line segment
  shown in the upper-left corner.
  The Fermi level and the Dirac point are shown schematically in each case.
  We also show for comparison the imaginary part of the self-energy for a conventional
  metal (dashed lines) (Ref.~\onlinecite{note:constant_g2N}).}
  \label{Figure_Im}
  \end{figure}

The electronic structure was described within the local density
approximation to density-functional
theory~\cite{ceperley_perdew}.
Valence electronic wavefunctions were expanded in a plane-waves basis~\cite{ihm:1979JPC_PW}
with a kinetic energy cutoff of 60~Ry. The  core-valence interaction was
treated by means of norm-conserving
pseudopotentials~\cite{troullier_fuchs}.
Lattice-dynamical properties were computed through density-functional perturbation theory
~\cite{baroni:2001RMP_DFPT}. We modeled an isolated graphene by a honeycomb lattice
of carbon atoms within a periodic supercell.
The graphene layers were separated by
$8.0$~\AA\ of vacuum~\cite{dubay:2003PRB_Graphene_Phonon},
and the relaxed C-C bond-length was 1.405~\AA.
Doped graphene was modeled
by varying the electronic density
and introducing
a neutralizing background charge.
We first calculated electronic and vibrational states and the associated
\EP\ matrix elements on $72\!\times\!72$ $\bk$-points
and $12\!\times\!12$ $\bq$-points in the BZ of graphene. 
Then, we determined the quantities needed to evaluate the self-energy given by
Eq.~(\ref{equation:selfenergy}) 
on a significantly finer grid of $1000\!\times\!1000$ $\bk$ and $\bk+\bq$ points
in the irreducible wedge of the BZ by using
a first-principles interpolation based on electron and phonon Wannier 
functions~\cite{giustino:2007PRL_Wannier,park:2007note_Wannier_detail,
park:2007note_Graphene_Phonon}.
The fine sampling of the BZ was found to be crucial
for convergence of the self-energy. 
In the calculation of the self-energy
we used a broadening parameter $\delta$ of 10~meV, comparable with the resolution
of state-of-the-art photoemission experiments~\cite{damascelli:2003RMP_ARPES}.
The calculations were performed with the electron and phonon
occupations~[Eq.~(\ref{equation:selfenergy})] corresponding to
$T\!=\!20$~K
to make connection with the
ARPES experiment~\cite{bostwick:2007NatPhys_Graphene}.
In what follows, we discuss the computed electron self-energy by focusing
on a straight segment perpendicular to the ${\rm \Gamma K}$ direction and centered
at the K point in the BZ [Fig.~\ref{Figure_Im}].

We note here that
within 2.5~eV from the Dirac point,
the angular dependence of the self-energy is insignificant
(at fixed energy $E$)~\cite{park:2007TBP_Graphene_angle}.
As a consequence, the \EP\ coupling parameter
$\lambda_n(\hat\bk)=-\left.{\partial\ 
\text{Re}\Sigma_{n\bk}(E)}/{\partial E}\right|_{E=E_{\rm F}}$
is isotropic in $\bk$-space.

Figure~\ref{Figure_Im} shows the calculated imaginary part of
the self-energy (which is closely related to the linewidth)
as a function of carrier energy, corresponding to three representative situations:
intrinsic, electron-doped,
and hole-doped graphene. We
here considered doping levels corresponding
to 4$\cdot10^{13}$~cm$^{-2}$ electrons or holes. The corresponding Fermi levels
were found to lie at $+0.64$~eV and $-0.66$~eV from the Dirac point, respectively.
In the intrinsic system, we found that the electron linewidth due to \EP\ interaction
is negligible ($<\!1$~meV) 
within an energy threshold $\hbar\omega_{\rm ph}$ for the emission of optical phonons
($\hbar\omega_{\rm ph}\approx0.2$~eV being a characteristic optical phonon frequency),
while it increases linearly beyond this threshold [Fig.~\ref{Figure_Im}(a)].
The scattering rate for electrons with energy below the optical phonon emission
threshold is negligible because (i) only optical phonons are effective in \EP\
scattering and (ii) Pauli's exclusion principle blocks transitions into occupied states.
On the other hand, the linear increase of the
linewidth above the optical phonon energy relates to the phase-space availability
for electronic transitions, and reflects the
linear variation of the density of states around the Dirac point in graphene.
The energy dependence of the electron linewidths in the 
electron-doped and the hole-doped systems
[Fig.~\ref{Figure_Im}(b) and Fig.~\ref{Figure_Im}(c), respectively] 
can be rationalized by a similar phase-space argument.
We denote by $E_{\rm D}$
the energy of the Dirac point with respect to the Fermi level.
For definiteness, we here consider the electron-doped situation ($E_{\rm D}<0$).
When the energy of the hole is exactly equal to $|E_{\rm D}|+\hbar\omega_{\rm ph}$
(i.e., at $-|E_{\rm D}|-\hbar\omega$ in Fig.~\ref{Figure_Im}),
there are no allowed final states for electronic transitions through optical
phonon emission, resulting in a vanishing scattering rate at zero temperature.
As the hole energy departs from $|E_{\rm D}|+\hbar\omega_{\rm ph}$,
the linewidth increases linearly, and exhibits a characteristic dip
at the Fermi level. The latter feature corresponds to forbidden phonon emission
processes, as discussed above for intrinsic graphene.
The calculated energy dependence of
the electron linewidth deviates substantially from the standard
result which applies to conventional
metals (Fig.~\ref{Figure_Im}, dashed line)~\cite{note:constant_g2N}.
The latter consists of a constant scattering rate above the phonon emission threshold, 
and fails in reproducing the 
features revealed by our {\it ab initio} calculations. 

  \begin{figure}
  \includegraphics[width=1.0\columnwidth]{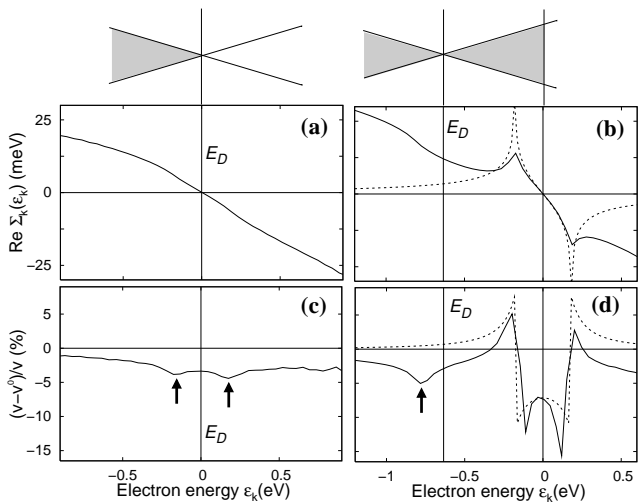}
  \caption{Calculated real part of the electron self-energy arising from
  the \EP\ interaction
 at $T=$ 20~K 
  (solid lines), for (a) intrinsic
  and (b) electron-doped graphene. The self-energy was evaluated along the
  reciprocal space line segment shown in Fig.\ \ref{Figure_Im}.
  The corresponding velocity renormalization $(v_{n\bk}-v^0_{n\bk})/v_{n\bk}$
  is shown in panels (c) and (d), respectively.
  We also report, for comparison,
  the real part of the self-energy and the velocity renormalization
  for a conventional metal (dashed lines) (Ref.~\onlinecite{note:constant_g2N}). 
  At variance with conventional metals, the group velocity in
  graphene shows additional dips when the carrier energy is
  $|E_{\rm D}|+\omega_{\rm ph}$ (arrows), reflecting the vanishing density 
  of states at the Dirac point.}
  \label{Figure_Re}
  \end{figure}

Figure~\ref{Figure_Re} shows the real part of the electron self-energy
arising from the \EP\ interaction, for intrinsic
and for electron-doped graphene.
The behavior of the hole-doped system
is qualitatively similar to the electron-doped case. 
While in conventional metals the real part of the self-energy 
decays at large hole energies ($E<-\hbar\omega_{\rm ph}$)
[Fig.~\ref{Figure_Re}(b), dashed line], 
the self-energy in graphene shows a monotonic increase in the same 
energy range [Fig.~\ref{Figure_Re}(b), solid line].
Since the wavevector dependence of the self-energy in graphene
within a few eV from the Fermi level is negligible
[i.e., $\Sigma_{n\bk}(E)\simeq\Sigma_n(E)$]~\cite{park:2007TBP_Graphene_angle},
we obtained the quasiparticle
strength $Z_{n\bk}= (1 - \partial\ \text{Re}\Sigma_{n\bk}/\partial E)^{-1}$
by evaluating $(1 - d\ \text{Re}\Sigma_{n\bk}(\varepsilon_\bk)/d\varepsilon_\bk)^{-1}$.
In all cases considered, the \EP\ interaction was found to reduce 
the non-interacting quasiparticle strength down to at most $Z_{n\bk}=0.93$
at the Fermi level.
This suggests that a quasiparticle picture is still
appropriate at low energy, the \EP\ interaction largely preserving 
the weakly perturbed Fermi-liquid behavior. 
The quasiparticle strength is related to the velocity
renormalization through $1-Z_{n\bk}^{-1} = (v_{n\bk}-v_{n\bk}^0)/v_{n\bk}$,
$v_{n\bk}^0$ and $v_{n\bk}$ being the non-interacting and the interacting 
velocity, respectively. The velocity renormalization is plotted in
Fig.~\ref{Figure_Re}(c) and Fig.~\ref{Figure_Re}(d)
for the intrinsic and the electron-doped system, respectively. 
The velocity renormalization at the Fermi level was found to increase with the doping
level, and amounts to $-4$~\%, $-8$~\%, and $-6$~\% in the intrinsic, the
electron-doped, and in the hole-doped system considered here.
Our results indicate that the velocity of Dirac fermions in graphene
is affected by the \EP\ interaction. This bears important
implications for the transport properties of graphene-based electronic devices. 

\begin{figure}
\includegraphics[width=1.0\columnwidth]{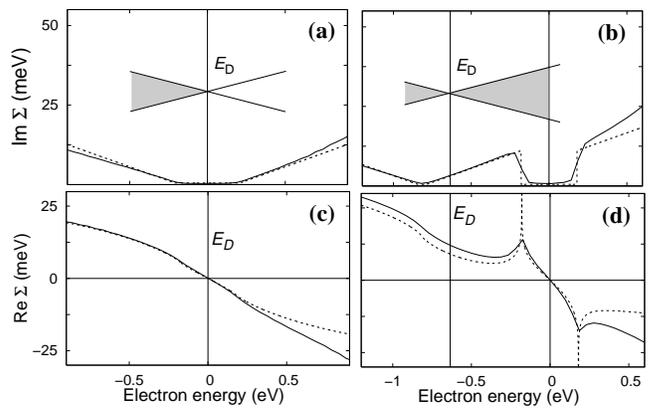}
\caption{Comparison between the electron self-energy obtained from
a first-principles calculation (solid lines) and a single-parameter
model (dashed lines):
imaginary (upper panels) and real (lower panels) part for
the intrinsic system (left) as well as for the electron-doped system (right).
Note that the horizontal energy ranges differ from those shown in
Fig.~\ref{Figure_Im}.}
\label{Figure_toy}
\end{figure}

In order to provide a simplified picture of the \EP\ interaction in graphene,
we analyzed the various \EP\ scattering processes contributing to
the electron lifetimes. We repeated our
calculations by restricting either the energy $\hbar\omega_{\bq\nu}$ or the
momentum transfer $\bq$ in Eq.~(\ref{equation:selfenergy}) to
limited ranges.
When only
the in-plane optical phonon modes between 174 and 204 meV
are taken into account in Eq.~(\ref{equation:selfenergy}), the electron 
linewidth is found to deviate from the full {\it ab initio} result by $15\%$ at most.
In contrast, when the momentum integration in Eq.~(\ref{equation:selfenergy})
is restricted to small regions around the high-symmetry points $\Gamma$ and K,
the linewidth is found to deviate significantly from the full calculation,
indicating that a proper account of the entire BZ is essential.
Based on this analysis, we devised a simplified single-parameter model
of the \EP\ interaction in graphene. We assumed: 
(i) linear electronic dispersions up to a few eV away from the
Dirac point~\cite{park:2007UNP_Graphene_cutoff},
(ii) an Einstein model
with the effective phonon energy $\hbar\omega_{\rm ph}$
set to that of the highest degenerate zone-center mode,
(iii) an effective \EP\ vertex $g$, independent of the electron and phonon
momenta~\cite{park:2007TBP_Graphene_angle}.
The \EP\ matrix element $g$ represents a free parameter in our simplified model,
and has been determined by matching the model self-energy with the full
{\it ab initio} result.
Within these assumptions, and with the Fermi level set to zero,
the imaginary part of the self-energy reads
\begin{equation}
{\rm Im}\ \Sigma(E)=
\frac{\sqrt{3}\,a^2}{16}\alpha_{\text{\tiny G}}^2 \, g^2 \,
\left|E-{\rm sgn}(E)\,\hbar\omega_{\rm ph}-E_{\rm D}\right|,
\label{equation:model}
\end{equation}
whenever $|E|$ exceeds the characteristic phonon energy $\hbar\omega_{\rm ph}$,
and vanishes otherwise.
In Eq.\ (\ref{equation:model}),
$a$ is the lattice parameter in Bohr units,
$\alpha_{\text{\tiny G}} = e^2/\hbar v^0 = 2.53$
is the  effective fine structure constant of graphene,
and $g$ is the average \EP\ matrix 
element in Rydberg units.
The fitting to our calculated {\it ab initio} self-energy gave $g=3.5 \cdot 10^{-2}$~Ry.
The real part of the model self-energy can be straightforwardly obtained
from Eq.~(\ref{equation:model}) through Kramers-Kr\"onig relations.
Figure~\ref{Figure_toy} shows that this simplified model
is in fairly good agreement with the full first-principles calculation.
Therefore, despite its simplicity, our single-parameter model captures
the qualitative features of the \EP\ interaction in graphene.

\begin{figure}
\includegraphics[width=0.8\columnwidth]{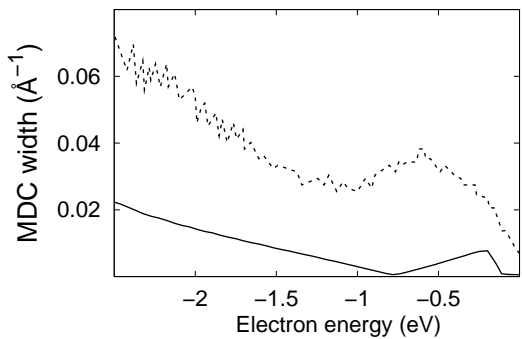}
\caption{Calculated width of the ARPES momentum distribution curve for electron
doped graphene (solid line) compared to the experimental result of
Ref.~\onlinecite{bostwick:2007NatPhys_Graphene}
(dashed line). In our calculation, the Fermi level was set in order to simulate 
the sample with $2.1 \cdot 10^{13}$ electrons/cm$^2$ in Fig.~3 of
Ref.~\onlinecite{bostwick:2007NatPhys_Graphene}.}
\label{Figure_MDCW}
\end{figure}

In Fig.~\ref{Figure_MDCW} we compare our first-principles calculations
with the width of the momentum distribution curve (MDC)
measured by ARPES experiments at 20~K
on graphene with a similar doping~\cite{bostwick:2007NatPhys_Graphene}.
The width $\Delta k_{n\bk}$ of the MDC
was calculated taking into account renormalization effects through
$\Delta k_{n\bk} = \ Z_{n\bk} \,
2\text{Im }\Sigma_{n\bk}/\hbar v_{n\bk}$~\cite{grimvall:1981_Metal_ElPh}.
Figure~\ref{Figure_MDCW} shows that, contrary to previous
findings~\cite{hwang:2006condmat_Graphene_GW},
the \EP\ interaction plays a significant role in reducing the carrier lifetime in
graphene, as it accounts for about a third
of the measured linewidth at large binding energies. 
The \EP\ contribution to the width of the MDC is found to increase linearly
at large binding energy, in agreement with experiment. 

In conclusion, we have computed from first-principles the velocity renormalization
and the carrier lifetimes in graphene arising from the \EP\ interaction and
we have reproduced these results with a simplified model.
The calculated energy-dependence of the phonon-induced electronic linewidths
is shown to originate from the linear electronic dispersions.
The renormalization of the Fermi velocity was found to be
$-4$~\% for intrinsic graphene and
$-8$~\% for an electron doping of $4\cdot 10^{13}$~cm$^{-2}$,
and is expected to affect the mobility of graphene-based
electronic and spintronic devices.

We thank Y.-W. Son, G. Samsonidze and M. Lazzeri for stimulating discussions.
We are grateful to M. Calandra and F. Mauri for carefully reading our manuscript
and for pointing out a numerical factor error in an earlier version.
This work was supported by NSF Grant
No. DMR04-39768 and by the Director, Office of Science, Office of Basic Energy
Sciences, Division of Materials Sciences and Engineering Division,
U.S. Department of Energy under Contract No. DE- AC02-05CH11231.
Computational resources have been provided by NPACI and NERSC.
Part of the calculations were performed using the
{\tt Quantum-Espresso}~\cite{baroni:2006_Espresso}
and {\tt Wannier}~\cite{mostofi:2006_Wannier} packages.

{\it Note added:} After submission of this manuscript we became aware of a related
work whose results are in agreement with our conclusions \cite{mauri}.


\begin{thebibliography}{10}
\bibitem{park:2007UNP_Graphene_cutoff}
We used an energy cutoff of 6 eV. The resulting self-energy was found to be
  largely insensitive to this choice.
\end{thebibliography}
\end{document}